\documentclass[a4paper,12pt]{article}
\usepackage{latexsym}
\usepackage{amsmath}
\usepackage{amssymb}
\usepackage{bm}
\usepackage{textcomp}
\usepackage{graphicx}
\usepackage[OT1]{fontenc}
\usepackage[english]{babel}

\usepackage{geometry}
\begin{document}

\linespread{1}

\begin{center}

\textbf{\large{R\'enyi entropy for particle systems as an instrument to enlarge the Boltzmannian concept of entropy: some holographic perspectives}}
\end{center}

\noindent
\begin{center}

Nicolò Masi

\begin{center}
INFN \& Alma Mater Studiorum - Bologna University,

Via Irnerio 46, I-40126 Bologna, Italy
\end{center}

\begin{center}
{\it Nicolo.Masi@bo.infn.it}
\end{center}

\end{center}

\begin{center}

\begin{abstract}
The R\'enyi entropy is a mathematical generalization of the concept of entropy and it encodes the total information of a system as a funtion of its order parameter $\alpha$. The meaning of the R\'enyi entropy in physics is not completely enstablished: here we determined a general and explicit representation of the R\'enyi entropy for whichever fluid of particles and spin-statistics, in the mechanical statistics framework. This allowed us to put physical constraints to the R\'enyi order $\alpha$, from main thermodynamical relations and entropy bounds of the holographic theories, defining how much we can enlarge the Boltmannian concept of entropy.
\end{abstract}
\end{center}

\small{\footnotesize{}}

The R\'enyi entropy [1-3] is a useful and powerful instrument to deal with a lot of problems in physics, statistics, mathematics, economy, genetics, biological systems and many others. It is defined as 
\begin{equation}
H_{\alpha}=\frac{1}{1-\alpha}ln(\sum p_{k}^{\alpha})
\label{1}
\end{equation}

R\'enyi called this quantity ''the measure of information
of order $\alpha$ associated with the probability distribution $P = (p_1,...p_n)$''. The order $\alpha$ is a positive real number $\in \ ]0,\infty[$, less $\alpha=1$. This general entropy was used in literature in the analysis of quantum entanglement [4-6] (computed also for Conformal Field Theories [7-9]), quantum correlations [10], spin systems and chains [11-13], and also to reformulate the Heisenberg uncertainty principle in terms of entropy [14]. In addition, it was applied to Black Holes (BH) [15] and holographic theories and related to the Ryu-Takayanagi formula [16,17]. In fact, the R\'enyi entropy has some peculiar features which must be underlined and that make it very versatile: it becomes the Boltzmannian entropy (or the Shannon's one for information theory and the Von Newmann's quantum entropy for entanglement treatment [2,17]) for $\alpha \rightarrow 1$, so $H_{\alpha \rightarrow 1}=S_{\text{th}}$. The R\'enyi entropy is a slowly decreasing function of $\alpha$; for $\alpha>\beta$ we have $H_{\alpha}\leq H_{\beta}$. Some inequalities in the other direction also hold, such as $H_2\leq 2H_{\infty }$. \\The R\'enyi entropy has not been completely integrated into standard thermodynamics as a generalized entropy. While many researchers have tried to modify statistical mechanics by changing the usual formula for entropy, so far the most convincing uses of R\'enyi entropy in physics seem to involve the limiting cases $\alpha\rightarrow0$ and $\alpha\rightarrow\infty$ [2,3]. However, it is not necessary to modify statistical mechanics to find a natural role for this entropy in physics, because it is closely related to the familiar concept of grand-potential and free energy, with the order $\alpha$ appearing as a ratio of temperatures.
In fact, it is well known in literature that R\'enyi entropy can be written in terms of grand-potenzial $\Omega$ [18-20], once defined two different temperatures
$T$ and  $T_0=T/\alpha $, that is an "alpha-shifted" temperature which carries the order of the associated R\'enyi entropy:
\begin{equation}
H_{\alpha }=- \frac{\Omega (T,\mu ) - \Omega \left(T_0,\mu\right)}{T-T_0}
\label{2}
\end{equation}

where $\mu$ is the chemical potential of the system. This is called an \textit{alpha-derivative} or \textit{alpha-deformation} of $\Omega$ [18] and it leads to a clear thermodynamical representation of $H_{\alpha }$, for a fixed unitary volume $V$, when $\alpha$ approches $1$ and $T_0=\frac{T}{\alpha }\overset{\alpha \rightarrow 1}{=}T$:
\begin{equation}
H_{\alpha }(T,\mu )\overset{\alpha \rightarrow 1}{=}-\frac{\partial \Omega (T,\mu )}{\partial T}=S_{\text{th}}(T,\mu )
\end{equation}

When we move to a canonical ensamble where $\mu\rightarrow0$, the grand-potenzial $\Omega$ can be substitued with the free energy of the system:
\begin{equation}
H_{\alpha }=- \frac{F(T) - F(T_0)}{T-T_0}
\label{3}
\end{equation}

As already shown with other methods and particular cases in [15,19,20], and from equation (3), it is not surprising that R\'enyi entropy is proportional to Boltzmann entropy through a function of $\alpha$ later on called $R_{\alpha }=R(\alpha)$: 
\begin{equation}
H_{\alpha}=R_{\alpha} S_{\text{th}}
\label{4}
\end{equation}

Here we want to demonstrate that the proportional coefficient between classical entropy and generalized entropy can be derived from statistical mechanics in a general way. In fact, using the thermodynamical formulas (2) and (4) of $H_{\alpha }$ described before, $R_{\alpha }=\frac{H_{\alpha }}{S_{\text{th}}}$ can be written as follows:
\begin{equation}
R_{\alpha }=\frac{1}{S_{\text{th}}}\cdot\frac{\Omega \left(T_0,\mu \right) - \Omega (T,\mu )}{T-T_0}\underset{\mu \rightarrow 0}{\overset{T_0=\frac{T}{\alpha }}{=}}\frac{1}{T S_{\text{th}}}\cdot\frac{\alpha }{\alpha -1}\cdot[F(T_0)-F(T)]
\label{5}
\end{equation}

where $F(T_0)=F(T /\alpha)$ is proportional to the associated alpha-shifted energy $E(T_0)$ of the system.\\ 
To procede with the calculation it is mandatory to define a general representation of the free energy in terms of energy and some universal thermodynamical coefficients we have already discussed in [21]. From now on we will use natural units. \\The fundamental Euler equation $S_{\text{th}}=\frac{E+pV}{T}-\ \frac{\mu N}{T}\overset{pV=wE}{=}\frac{E(1+w)-\mu N}{T}$ [22] (rewritten using the equation of state formalism), that links entropy, energy, pressure $p$ and the chemical potential $\mu$ of the $N$ elements of the system of volume $V$, is the starting point of the examination: when $\mu \rightarrow 0$, the entropy to energy ratio ${S_{\text{th}}}/{E}$ assumes the form  
\begin{equation}
\frac{S_{\text{th}}}{E}=\frac{1+w}{T}=(\frac{4}{3}, \frac{5}{3})\cdot \frac{1}{T} 
\label{6} 
\end{equation}

$w$ is the coefficient of the equation of state (EOS) of the energy-matter gas and, as well-known in literature, it is worth ${1}/{3}$ for photons and low-pressure ultrarelativistic particles, and $2/3$ for massive pressure-carriers particles. Consequently, $4/3$ represents the massless radiation case and $5/3$ the massive one [21]. We generically call $l$ the inverse of this proportionality coefficient between the ratio ${S_{\text{th}}}/{E}$ and the inverse of the temperature $T$:
\begin{equation}
S_{\text{th}}=l^{-1} \frac{E}{T}, 
\label{7}
\end{equation}

and
\begin{equation}
\frac{S_{\text{th}}}{E}=\frac{1+w}{T}=(lT)^{-1}\geq T^{-1}
\label{8}
\end{equation}

The special cases $l=\frac{3}{4},\frac{3}{5}$ stand for the above mentioned thermodynamical ensambles.
It is quite trivial to note that $l^{-1}$ is always greater than 1 (or equal) for all ''good fluids'' with $p/\rho\geq0$.
The relation in (9) can be generalized using the previously defined R\'enyi entropy and the alpha-shifted temperature as follows:
\begin{equation}
\frac{H_{\alpha }}{E(T_0)}\geq T_0^{-1} 
\label{9}
\end{equation}

This defines an enlarged relation that is a function of the R\'enyi order $\alpha$ and the statistical definition of $E$. The apparently accidental observation that the ratio cannot be greater than $T_0^{-1}$ will be strongly connected with the inability of the system of reaching the ultimate space-time scale ${\lambda }^*={1}/{(\pi T})$, as stressed and largely discussed in [21][23-25]: $l^{-1}$ and its alpha-generalization we are going to compute cannot be less than 1, to preserve this fundamental bound. We will come back later on this issue. \\
Let's define the most general integral representation of energy $E(T)$ and free energy $F(T)$ [22][26-28] (for unitary volume and alpha-shifted temperature), leaving the statistics phase-space exponent $n$, \textit{i.e.} the exponent of the energy at the numerator which comes from the ordinary number of states in the one-particle phase space, free to variate:
\begin{equation}
E(T_0)=\int^{\infty }_0{\nu N\left(\nu \right)}d\nu =\frac{g}{2^{n/3} \pi ^2 M^{n-3}} \int \frac{\nu ^n}{e^{\frac{\nu }{T/\alpha}}\pm 1} \, d\nu = \\
\label{10}
\end{equation}

\[ =\left\{ \begin{array}{c}
\frac{g \ 2^{-\frac{n}{3}} M^{3-n} \text{Li}_{n+1}(1) \Gamma (n+1) \left(\frac{\alpha }{T}\right)^{-n-1}}{\pi ^2}\ ,\ \ Bosons  \\ 
\frac{g \ 2^{-\frac{4 n}{3}} \left(2^n-1\right) M^{3-n} \zeta (n+1) \Gamma (n+1) \left(\frac{\alpha }{T}\right)^{-n-1}}{\pi ^2}\ ,\ \ Fermions \end{array}
 \right.\]

\begin{equation}
F(T_0)=T_0 \frac{g}{2^{n/3} \pi ^2 M^{n-3}} \int \nu ^{n-1} ln(1\pm e^{-\frac{\nu }{T/\alpha})}\, d\nu =\pm w \cdot \frac{g}{2^{n/3} \pi ^2 M^{n-3}}  \int \frac{\nu ^n}{e^{\frac{\nu }{T/\alpha}}\pm 1} \, d\nu 
\label{11}
\end{equation}

where $g$ is the spin-statistics factor, $\nu $ is the energy of the particle, $N\left(\nu \right)$ is the usual particle population as a function of energy, $M$ the total mass of the system and $n$, as mentioned above, the statistics phase-space exponent, which is 3 for massless particles (photons and ultrarelativistic particles, recovering Planck formula for massless radiation) and 3/2 for massive particles (according to Fermi-Dirac statistics in three spatial dimensions). ${\rm Li}_{n+1}$ is the polylogarithm, $\zeta (n+1)$ the Riemann Zeta and $\Gamma (n+1)$ the Euler Gamma function. The sign $+$ at the denominator stands for fermions, the sign $-$ for bosons. Here we have assumed that $\mu \equiv 0$, that is a solid condition for massless bosons and also for ultrarelativistic particles [21,22]; this is further supported by the fact that we want to use these thermodynamical relations for non-open systems. It must be emphasized that this thermodynamic description is consistent only if the system is large enough, that means it is valid in the regime of short wavelenghts ${\lambda }$ w.r.t. the spatial scale of the system. The physical reason for the different exponents, i.e. 3/2 versus 3, lies in the different one-particle density of states for fermions, which comes from the classical phase space (and from the number of spatial dimensions), and for bosons, which follow a Planck black body law; as for the spin, fermions have a fractional value for the temperature power, whereas bosons have an integer one.\\The free energy is computed as the natural logarithm of the grand canonical partition function in the $\mu\rightarrow 0$ limit [22]. 

The parameter $w$, which becomes the ratio of this two physical quantities in (11) and (12) [22], is independent from the sign $\pm$ of the statistical factor, from the choice of the statistic phase-space exponent and from the R\'enyi order $\alpha$: it is a universal constant which is always defined for $n\geq1$. It defines the "distance" between the energy and the canonical free energy:
\begin{equation}
F(T)=-w E(T)= -(l^{-1}-1) E(T)
\label{12}
\end{equation}

Now we can construct a consistent derivation of $R_{\alpha}$ from equation (6), using the previous formulas:
\begin{equation}
R_{\alpha }=(T S_{\text{th}})^{-1} (l^{-1}-1) \frac{\alpha }{\alpha -1}\cdot [E(T)-E(T_0)]\overset{T S_{\text{th}}=\frac{E(T)}{l}}{=}(1-l) \frac{\alpha }{\alpha -1}\left(1-\frac{E(T_0)}{E(T)}\right)
\label{13}
\end{equation}

where we used Eq. (8) for the last equality. The resulting R\'enyi coefficient is a function of $\alpha$, $l$ (and therefore $w$) and of the ratio $\frac{E(T_0)}{E(T)}$ between the two different energies, the alpha-shifted in $T_o$ and the Boltmannian one in $T$; this ratio does not depend on the particle statistics $\pm$ sign and is worth $\alpha^{-(n+1)}$.  From (14) an explicit formula for the R\'enyi entropy of a whatever particle system can be achieved:
\begin{equation}
H_{\alpha }=\frac{(1-l)}{l} \frac{\alpha}{(\alpha -1)}\cdot \frac{[E(T)-E(T_0)]}{T}=w \frac{\alpha}{(\alpha -1)}\cdot \frac{[E(T)-E(T_0)]}{T}
\label{14}
\end{equation}

This is one of the main results of the dissertation.\\ 
If one wants to correlate all the "special coefficients" of the thermodynamical description, it can be noted that, transforming the first free energy integral in (12) using integration per part [22], the resulting $w$ factor is a simple function of $n$:
\begin{equation}
w= n^{-1}=l^{-1}-1
\label{16}
\end{equation}

For statistic phase-space exponent $n=3,3/2$, the well-known results for radiation fluids and pressure-matter fluids are immediately achieved, respectively. Because of Eq. (16), any condition in one of these coefficients can be translated into another for a different quantity. For example, it can be noted that, for an equation of state $p=w \rho$, the $l^{-1}\geq1$ condition corresponds to the so-called in cosmology "Zel'dovich range" $0\leq w \leq 1$ [29], where $w=1$ only when $l=1/2$ and $n=1$, \textit{i.e.} when the number density $N(\nu)$ is a constant: $w=1$ ($p = \rho$) represents also the causal limit where the fluid's speed of sound is equal to the speed of light. \\ 

This reasoning can be also applied to the holographic sector [30]. The holographic theory includes some fundamental bounds: the Universal Entropy Bound by Bekenstein (UEB) [31-33] and the Holographic one [34,35], generalized by Bousso in his light-sheet reformulation [30]. The behavior of the bounds once applied to thermodynamic particle ensembles has been already explored in [21,36]. Even if they deal, in principle, with different physical quantities - energy and geometrical surface - for different systems, they give complementary information of the quantum-gravity properties of extreme systems. They are linked to a third bound, the Time Times Temperature bound by Hod [37,38], which is strongly but not explicitely interconnected. See [21][30] for details and insights of the holographic bounds. \\ 
One of the fundamental results we obtained in [21] is that, from the UEB bound by Bekenstein, a temporal constraint ca be put: \textit{i.e.} $\tau \ge \frac{1+w}{\pi T}$. Considering Eq. (16), the impossibility to reach and overcome the ${1}/{(\pi T})$ quantum can be easily interpreted as a clear constraint in $w-l-n$, that is $w\geq0$. In this sense the time inequality $\tau \ge \frac{1+w}{\pi T}$ suggests that also the TTT bound by Hod, that is $\tau_{relax} \ge \frac{1}{\pi T}$, corresponds to the null-positivity condition of the EOS coefficient $w$. This is an important hint of the meaning correspondence between mechanical statistics quantities and holography bounds. Hereafter this relation will be examined.\\ 

In addition, it must be noted that, using relation (16) between $w$ and $n$, the $R_{\alpha}$ coefficient in (14) can be easily written as a function of $w$ and $\alpha$:
\begin{equation}
R_{\alpha}=\frac{w}{1+w}\cdot\frac{\alpha^{\frac{1+w}{w}}-1}{\alpha^{1/w}(\alpha-1)}
\label{17}
\end{equation}

Now we have all the elements to explicitely compute the R\'enyi coefficients for bosons and fermions, respectively:
\begin{equation}
R_{\alpha }^{\text{Bose}}=\frac{1}{4} \left(\frac{1}{\alpha ^3}+\frac{1}{\alpha ^2}+\frac{1}{\alpha }+1\right)=(1-l) \sum _{i=0}^3 \alpha ^{-i}=\frac{w}{1+w} \sum _{i=0}^{1/w} \alpha ^{-i}
\label{18}
\end{equation}

\begin{equation}
R_{\alpha }^{\text{Fermi}}=(1-l) \frac{\left(\alpha ^{3/2}+\alpha ^2+\sqrt{\alpha }+\alpha +1\right) }{\alpha ^{3/2}+\alpha ^2}
\label{19}
\end{equation}

Bose coefficient is a sum of integer powers of $\alpha$, up to $n=3$, whereas the Fermi one is described by a sum of half-integer powers up to $n=3/2$. As previously mentioned, the result (18) for the boson case has been already achieved in [15,20], with different instruments but in a less comprehensive and general way, without any explicit link between $R_{\alpha}$ and the intrinsic mechanical statistics behaviour of particle gases encoded in $w$. Moreover, formula (17) stands for all thermodynamical ensambles in which $\mu\rightarrow0$ or $\frac{\left|\mu n\right|}{\rho }\ll \ 1$, \textit{i.e.} for "enough kinetic" systems (see [21] for details). \\
The ratios between the two bosonic and fermionic R\'enyi alpha-entropies we can obtain using (15), for massive $w=2/3$ and massless $w=1/3$ cases respectively, manifest some peculiar features:
\begin{eqnarray}
\frac{H_{\alpha }^{\text{Fermi}}}{H_{\alpha }^{\text{Bose}}}=\frac{13}{20}>\frac{1}{2},\  M\neq 0 
\\ \medskip
\frac{H_{\alpha }^{\text{Fermi}}}{H_{\alpha }^{\text{Bose}}}=\frac{7}{16}<\frac{1}{2},\  M=0
\label{20}
\end{eqnarray}

For gases of massive pressure-carriers particles the Fermi to Bose ratio is a fraction greater than $1/2$, whereas for radiation or ultrarelativistic particles ensambles the R\'enyi entropy of fermonions is always less than half the bosons entropy. We already know that fermionic entropy and energy is nevertheless less than the corresponding bosonic one and this continues to be true for R\'enyi alpha-entropy, which shows in (20,21) constant values that are independent from $\alpha$.\\
For what concerns the hierarchy between different alpha orders of the thus obtained Rènyi entropies around $\alpha=1$, \textit{i.e.} if one wants to calculate how much the R\'enyi entropy is distant from the Boltzmann entropy, the following examples stand for bosons and fermions, for $\alpha=1/2(<1)$ and $\alpha=2(>1)$:  
\begin{eqnarray}
H_{\frac{1}{2}}^{\text{Bose}}=\frac{15}{4} H_1^{\text{Bose}}\simeq 4 S_{\text{th}} \\
H_2^{\text{Bose}}=\frac{15}{32} H_1^{\text{Bose}}\simeq \frac{S_{\text{th}}}{2}, 
\label{21}
\end{eqnarray}

\begin{eqnarray}
H_{\frac{1}{2}}^{\text{Fermi}}\simeq 2 S_{\text{th}} \\
H_2^{\text{Fermi}}\simeq \frac{2 S_{\text{th}}}{3}
\label{22} 
\end{eqnarray}

R\'enyi entropy for bosons deviate from the classical one faster than the one for fermions, reaching a value about four times greater for $\alpha=1/2$, and a 
halved value for $\alpha=2$. Another important relation than has to be stressed is the distance between the classical $\alpha\rightarrow1$ case and the so-called
\textit{Min Entropy}, that corresponds to $H_{\infty}$, is:
\begin{equation}
H_{\alpha\rightarrow 1}-H_{\alpha\rightarrow \infty}=l S_{\text{th}}=(w+1)^{-1} \frac{E}{T}
\label{23}
\end{equation}

The difference between Boltzmann entropy and the smallest conceivable entropy $H_{\alpha\rightarrow \infty}$ is \textit{l-times} the Boltzman entropy itself, where $l$ should be always less than 1 for the previous considerations. This is another way to interpret the special number $l$ which emerges from this mechanical statistics analysis. The important Equation (26) stands for all types of statistics: both for fermions and bosons.\\
These general results so far obtained can be used in order to determine physical constraints for the R\'enyi alpha-order and therefore for the extension of the Boltmannian entropy concept. 
First of all we start with the enlarged thermodynamical relation (10), where explicit values for the R\'enyi entropy, for the alpha-shifted energy and the scaled temperature $T_0$ can be inserted:
\begin{equation}
\frac{H_{\alpha }}{E(T_0)}=\frac{1}{T}\cdot\frac{l^{-1} - 1}{\alpha^{-1} - 1}\cdot(1-\frac{E(T)}{E(T_0)})=\frac{1}{T}\cdot w\cdot\alpha \frac{\alpha ^{\frac{1+w}{w}}-1}{\alpha -1}\overset{\alpha \rightarrow 1}{\longrightarrow }\frac{1+w}{T}\geq \frac{1}{T}
\label{24}
\end{equation}

When $\alpha\rightarrow1$ the Boltzmannian case is recovered. The ratio $H_{\alpha }/E(T_0)=f(w,\alpha)T^{-1}$ (we called $f$ the analytic coefficient of $T^{-1}$ which generalizes $l^{-1}$) does not depend on the fermionic or bosonic nature of the system, but only on $w$ and $\alpha$. Then we want to fix the fluid equation of state parameter $w$ and study when the quantity in (27) is equal to $T^{-1}$, \textit{i.e.} when $f(w,\alpha)=1$ and the entropy-energy ratio reaches its minimum value for $w\geq 0$ fluids:
\begin{equation}
w\cdot\alpha \frac{\alpha ^{\frac{1+w}{w}}-1}{\alpha -1}=1
\label{25}
\end{equation}

Solving this equation for $\alpha$, for the massless/radiation case $w=1/3$ and the massive one $w=2/3$, it is obtained:
\begin{eqnarray}
w= 1/3: \ \alpha^* \simeq 0.88818 \\
w= 2/3: \ \alpha^{**} \simeq 0.73835
\label{26}
\end{eqnarray}

The massive fluid shows a lower $\alpha$ and a higher content of entropy and, consequently, it gives a less stringent constraint than a massless bosonic system.
It follows that the difference between the generalized R\'enyi entropy and the Boltzmann one has an upper limit defined by the alpha-orders in (29),(30):
\begin{eqnarray}
H_{\alpha ^*}^{\text{massless}}-S_{\text{th}}=S_{\text{th}}\left(R_{\alpha ^*}^{\text{massless}}-1\right)\simeq 0.2052 S_{\text{th}} \ \Rightarrow \Delta_{H_{\alpha^* }^{\text{massless}},S_{\text{th}}}\leq 0.2052 S_{\text{th}} \\
H_{\alpha ^{**}}^{\text{massive}}-S_{\text{th}}=S_{\text{th}}\left(R_{\alpha ^{**}}^{\text{massive}}-1\right)\simeq 0.2809 S_{\text{th}} \ \Rightarrow \Delta_{H_{\alpha^{**} }^{\text{massive}},S_{\text{th}}}\leq 0.2809 S_{\text{th}}
\label{27}
\end{eqnarray}

For the radiation case, as expected, the result is much stringent and constrains $\alpha$ to be greater than about 0.89 in order to fulfill the thermodynamical relation (27): this is the \textit{maximal entropy description} of the system which satisfies (28). In this sense, this is a physical bound for the alpha-order of the generalized R\'enyi entropy, which cannot assume arbitrary values and cannot enlarge the Boltzmannian entropy description beyond a certain limit. If relation (28) is solved for $w\rightarrow1$, \textit{i.e} for a stiff fluid, the solution in alpha is $\alpha\simeq 0.61803$, which represents the less stringent case and the maximum conceivable budget of entropy under the thermodynamical assumption (27).\\
These quantitative prediction for the R\'enyi alpha-order can be achieved also from the holographic bounds, as we can argue from simple considerations. The Bekenstein's bound [31-33] can be rewritten in natural units in terms of alpha-shifted energy and R\'enyi entropy in the following way:
\begin{equation}
\noindent 
S\le 2\pi R E \Rightarrow H_{\alpha}\le 2\pi R E(T_0) 
\label{28}
\end{equation}

If we want to compare it with relation (27), we see that $H_{\alpha}/E(T_0)\leq 2\pi R $, \textit{i.e.} the entropy-energy ratio has an upper limit as a function of the radius $R$ of the system. Using the general formula (15) it follows:
\begin{equation}
\frac{H_{\alpha}}{\pi E(T_0)}=\frac{1}{\pi T}\cdot\frac{l^{-1} - 1}{\alpha^{-1} - 1}\cdot(1-\frac{E(T)}{E(T_0)})=\frac{1}{\pi T}\cdot w\cdot\alpha \frac{\alpha ^{\frac{1+w}{w}}-1}{\alpha -1}\leq 2 R
\label{29}
\end{equation}
 
It must be remembered that the Universal Energy Bound is always relevant for weakly
self-gravitating isolated systems with spherical symmetry (and for gravitating systems in
general, once defined the physical quantities in a significant way), and for these systems
it is a much stronger bound than the holographic one [21,30].\\ 
To put in evidence the meaning of this UEB bound representation, we can link it to a Schwarschild black hole with event horizon radius $R$: for Schwarschild BHs the well-known relation $2R=\frac{1}{2}\cdot \frac{1}{\pi T}$ stands [39-41].
Hence, from Bekenstein's bound plus radius-temperature Schwarschild black hole condition for anyone statistics (and type of fluids) we obtain:
\begin{equation}
w\cdot\alpha \frac{\alpha ^{\frac{1+w}{w}}-1}{\alpha -1}=1/2, \ 
\label{30}
\end{equation}

Fixing $w$ for the radiation case, that is always the most extreme physical case we know and we can consistently put in the equation, a peculiar estimate for $\alpha$ is achieved:
\begin{equation}
 w= \frac{1}{3}\Rightarrow\ \alpha_{UEB+BH}\simeq 0.64447
 \label{31}
\end{equation} 

Here we used $w=1/3$ following the reasoning in [15], where the authors show that the R\'enyi entropy for a Schwarschild black hole is the same as the photon gas, if defined on the event horizon. We have obtained another lower restriction for $\alpha$ (when $f(w,\alpha)=1/2$) and, consequently, an upper bound for R\'enyi entropy; so $\alpha$ must be $\geq 0.64447$ in order to fulfill the Bekenstein-BH condition: such a big entropy content can be reached only by extremal objects, such as black holes. Absurdly, if we use lower $w$ (1/4, 1/5, etc...), we obtain increasing and more constraining values of $\alpha_{UEB-BH}$ (0.70901, 0.75380, etc, respectively).\\ 
If we want to add an additional and noteworthy constraint to the R\'enyi entropy starting from the UEB and recover the fundamental condition (28), the TTT bound [37,38] can be exploited for this aim. In fact, the minimum necessary time to transfer matter-energy through the sphere diameter of a spherical physical system is ${\tau }_{light}=\frac{2R}{c}$ and therefore the general relations $\tau \ge 2R$ and $\tau >2R\ $stand, respectively, for massless and massive particles. So, applying $\tau \ge 2R $ to (34), the bound implies that:
\begin{equation}
\frac{1}{\pi T}\cdot w\cdot\alpha \frac{\alpha^{\frac{1+w}{w}}-1}{\alpha -1}\leq \tau
\label{32}
\end{equation}

and, to fulfill the TTT equivalence $\tau=1/(\pi T)$,
\begin{equation}
w\cdot\alpha \frac{\alpha ^{\frac{1+w}{w}}-1}{\alpha -1}=1 
\label{33}
\end{equation}

We obtained again equation (28), but using an "holographic prospect", for a system which satisfies and saturates the UEB and the TTT bound at the same time. \\

Concluding, we have developed a consistent treatment of the R\'enyi entropy in the mechanical statistics context, giving explicit representations for particle systems. This has been done for whichever statistics, type of fluid and R\'enyi order alpha: it allows an enlarged thermodynamical description as a function of two parameters $\alpha$ and $w$. This also grants an insight of what a generalized thermodynamics can be, regardless the peculiar Boltzmannian case $\alpha\rightarrow1$ and the standard massless/massive gases $w=1/3, 2/3$. \\ To test how much the classical thermal entropy can be extended, the R\'enyi entropy has been analyzed: first we obtained a general constraint $\alpha\geq 0.88818$ for a \textit{maximal entropy description} of a thermodynamical system. It subtends a null-positive condition for pressure and energy density in the EOS of the fluid. Then we added Bekenstein's prescription along with TTT bound, achieving the same constraining value for the R\'enyi entropy. \\ As stressed in [21], the TTT and UEB bounds encode in a complementary way the microscopic behaviour of the space-time towards the matter-energy content, through the $1/(\pi T)$ quantum: so the constraints from (28)/(38) give an estimate of the maximum budget of entropy that is stored on a thermodynamically consistent holographic space-time, once enlarged the Boltzmannian treatment.\\

\textsl{In deferential memory of Jacob Bekenstein.}
 
\noindent 
\begin{center}
\large{\textbf{References}}
\end{center}

\begin{enumerate}

\item A. R\'enyi, \textit{On measures of information and entropy}, Proceedings of the fourth Berkeley Symposium on Mathematics, Statistics and Probability 1960, pp. 547561, 1961
\item P. A. Bromiley, N. A. Thacker, E. Bouhova-Thacker, \textit{Shannon Entropy, Renyi Entropy, and Information}, Statistics and Segmentation Series (2008-001), 2004.
\item E. Beadle et al., Signals, Systems and Computers, 42nd Asilomar Conference, 2008

\item F. A. Bovino, G. Castagnoliet al., Phys. Rev. Lett. \textbf{95}, 240407 (2005).
\item I. Bengtsson and K. Zyczkowski, \textit{Geometry of Quantum States}, Cambridge University Press, 2006.
\item L. Wang and M. Troyer, Phys. Rev. Lett. \textbf{113}, 110401 (2014).

\item D. V. Fursaev, arXiv:1201.1702, 2012
\item J. Hung et al., arXiv:1110.1084, 2011.
\item I. R. Klebanov et al., arXiv:1111.6290, 2012.

\item P. L\'evay, S. Nagy, J. Pipek, Phys. Rev. A \textbf{72}, 022302 (2005).

\item S. Gnutzmann and K. Zyczkowski, J. Phys. A: Math. Gen. \textbf{34}, 10123 (2001).
\item F. Verstraete and J. I. Cirac, Phys. Rev. B \textbf{73}, 094423 (2006).
\item Franchini, F. et al., J.Phys. A \textbf{41} (2008) 025302.

\item I. Bialynicki-Birula, Phys. Rev. A \textbf{74}, 052101 (2006).

\item A. Bialas and W. Czyz, arXiv:0801.4645, 2008

\item S. Ryu and T. Takayanagi, JHEP \textbf{08} (2006) 045.

\item M. Headrick, Phys.Rev. D \textbf{82} (2010) 126010.

\item J. C. Baez, arXiv:1102.2098, 2011.
\item A. Bialas and W. Czyz, Acta Phys. Pol. B \textbf{31} (2000) 2803
\item A. Bialas, W. Czyz, K. Zalewski, arXiv:hep-ph/0606319, 2006. 

\item N. Masi, Int. J. Mod. Phys. D \textbf{24}, No. 10 (2015) 1550076.

\item  W. Greiner et al., \textit{Thermodynamics And Statistical Mechanics}, Springer, 1997.

\item  A. Pesci, arXiv:1108.5066v1, 2011.
\item   A. Pesci, arXiv:0803.2642v2, 2008
\item   A. Pesci, arXiv:0807.0300v2, 2008.

\item  D. A. McQuarrie, \textit{Statistical Mechanics}, University Science Books, 2000.
\item  K. Huang, \textit{Statistical Mechanics}, John Wiley and Sons, 1987.
\item  B. H. Lavenda, \textit{A New Perspective of Thermodynamics}, Springer 2010.

\item P. Coles, F. Lucchin, \textit{Cosmology: the origin and evolution of cosmic structure},Wiley (2ed.), 2002.

\item  R. Bousso, arXiv:hep-th/0203101v2.

\item  J. D. Bekenstein, arXiv:quant-ph/0404042v1.
\item  J. D. Bekenstein, Phys. Rev. D \textbf{7}, No. 8, 1973.
\item  J. D. Bekenstein, Phys. Rev. D \textbf{23}, No. 2, 1981.

\item   G. 't Hooft, \textit{Dimensional reduction in quantum gravity}, in Salam--festschrifft, A. Aly, J. Ellis, and  S. Randjbar--Daemi, eds. (World Scientific, Singapore, 1993), arXiv gr--qc/9310026.
\item   L. Susskind, J. Math. Phys. \textbf{36}, 6377 (1995).

\item   N. Masi, arXiv:1109.4384.

\item  S. Hod, Phys. Rev. D \textbf{75 }(2007) 064013.
\item  S. Hod, Phys. Rev. D \textbf{78}, 084035 (2008).

\item   T. Padmanabhan, \textit{Gravitation -- Foundations and Frontiers}, Cambridge University Press, 2010.
\item   R. Wald, \textit{General Relativity}, University of Chicago Press, 1984.
\item   X. Calmet, \textit{Quantum Aspects of Black holes}, Springer, 2015.

\end{enumerate}

\noindent

\end{document}